\documentstyle[eqsecnum,twocolumn,floats,aps]{revtex}
\input{epsf}

\begin{document}
\draft
\wideabs{
\preprint{HEP/123-qed}
\title{Microwave Spectroscopy of Thermally Excited Quasiparticles
in $\bf YBa_2Cu_3O_{6.99}$}
\author{A. Hosseini, R. Harris, Saeid Kamal, P. Dosanjh,
J. Preston, Ruixing Liang, W.N. Hardy and D.A. Bonn}
\address{Dept. of Physics and Astronomy, University of British Columbia,
6224 Agricultural Rd., Vancouver, BC, V6T 1Z1, Canada}
\date{\today}
\maketitle
\begin{abstract}
We present here the microwave surface impedance of
a high purity crystal of $YBa_2Cu_3O_{6.99}$ measured
at 5 frequencies between 1 and 75 GHz. This data set reveals the
main features of the conductivity spectrum of the thermally excited
quasiparticles in the superconducting state.
Below 20 K there is a regime of extremely long quasiparticle lifetimes,
due to both the collapse of inelastic
scattering below $T_c$ and the very weak impurity scattering in the
high purity 
$BaZrO_3$-grown crystal used in this study. Above 20 K, the scattering
increases dramatically, initially at least as fast as $T^4$.

\end{abstract}
}

\narrowtext

\section{Introduction}
\label{sec:introduction}

Over the past few years measurements of electrodynamic properties at
microwave frequencies have proven to be a particularly fruitful technique
for studying the superconducting state of the high temperature superconductors.
A key strength of the technique is that measurements of the real and
imaginary part of the surface impedance $Z_s(\omega,T)$ provide complementary 
information on two aspects of the superconducting state; the superfluid density
and the low energy excitations out
of the condensate. 

Measurements of the imaginary part of the surface impedance
$X_s(T)$ provide a direct measurement of the penetration depth $\lambda(T)$ and
thus give a direct measure of the temperature dependence of the superfluid 
density. Such measurements have produced a wealth of information on the
superconducting state. The widespread observation of a linear temperature
dependence of $\lambda(T)$ at low T in many of the superconducting cuprates
\cite{hardy93,broun97,devaulchier96,lee96,jacobs95}
has been a key piece of evidence suggesting nodes in the energy gap in these
materials. Near $T_c$ the temperature dependence of the superfluid density has
provided evidence of 3DXY critical fluctuations over a wide temperature range
in $YBa_2Cu_3O_{7-\delta}$ \cite{kamal94}. 

Measurements of the real part
of the surface impedance $R_s(T)$, when combined with the measurements of
$\lambda(T)$, can be used to determine the real part of the microwave
conductivity $\sigma_1(T)$,
which is essentially electromagnetic absorbtion by quasiparticles excited
out of the condensate (either thermally excited quasiparticles or
excitations created by the absorbtion of photons).
Early measurements of $R_s(T)$ at a few 
GHz exhibited a broad peak below $T_c$, caused by a very large peak in 
$\sigma_1(T)$\cite{bonn92} at these low frequencies.
This peak was also observed at THz frequencies by Nuss et al.
\cite{nuss91}, and was attributed to a competition between 
two temperature dependences; the overall decrease with temperature of the
number of thermally excited quasiparticles competing with a rapid increase 
below $T_c$ of the transport lifetime of these quasiparticles. This rapid 
increase in lifetime has been interpreted as a collapse of the inelastic 
scattering processes responsible for the large normal state resistivity of
the high temperature superconductors and is an effect that is observed in
thermal conductivity measurements \cite{yu92}
as well as in other electromagnetic
absorption measurements at microwave 
\cite{broun97,shibauchi92,mao95,zhang94,anlage94,shibauchi94,kitano95,dahne95}, 
far infrared\cite{romero92} and THz frequencies
\cite{nuss91,spielman94,pimenov98}.
The rapid increase
in quasiparticle lifetime is well established by these measurements and the very
long quasiparticle mean free paths resulting from this have been 
corroborated further by the thermal Hall effect measurements of Krishana et al.
\cite{krishana95}.
However, obtaining
a quantitative determination of the scattering lifetime and the details
of its temperature dependence has been hampered by the need to use models
to interpret the existing microwave data. The problem has been that although
the step from the $R_s(T)$ and $\lambda(T)$ to the conductivity $\sigma_1(T)$
is a matter of straightforward 
superconductor electrodynamics when in the local limit, the extraction of
a scattering time from $\sigma_1(T)$ suffered from a dependence on an 
assumed model for the shape of the quasiparticle conductivity spectrum
$\sigma_1(\omega)$. 

In this paper we present measurements at 5 different microwave frequencies,
giving enough spectroscopic detail over a wide enough frequency range
to produce a rather complete picture of the evolution of $\sigma_1(\omega,T)$
in the superconducting state. These results support the early model based on 
an ansatz that the thermally excited quasiparticles have a nearly Drude-shaped
conductivity spectrum and now provide a much clearer measurement of the
temperature dependence of the quasiparticle scattering
rate below $T_c$. We find that
this scattering rate becomes extremely small and appears to become
essentially
temperature independent below about
20 K. At higher temperatures the scattering rate
increases rapidly, initially at least as fast as $T^4$.
In the 
following section we will describe the techniques used to produce this 
information and in Section \ref{sec:results} we will present the results
along with the details of the extraction of the microwave conductivity from
the surface impedance measurements. In Section \ref{sec:tau} 
the microwave conductivity spectra will be presented
along with the results of fits
that generate the temperature dependence of the quasiparticle lifetime.
In Section \ref{sec:discussion}
we will point out some of
the implications of these results and in
particular compare them to the present literature on
conductivity in a $d$-wave superconductor.

\section{Experimental Techniques}
\label{sec:experiment}

The property being directly probed in a microwave measurement on a 
superconductor is the complex surface impedance $Z_s(T)\, = \, R_s(T)+iX_s(T)$.
One simplifying feature of the high temperature superconductors is that the
superfluid response falls in the limit of local electrodynamics, so that
$R_s(T)$ and $X_s(T)$ are related to the complex conductivity 
$\sigma(\omega, T)\, = \, \sigma_1(\omega,T) - i\sigma_2(\omega,T)$ in a 
straightforward way. This simple limit arises from the very small coherence
length in these materials, which guarantees that $\xi \ll \lambda$. 
A particularly
simple regime of the local limit occurs when $\sigma_2\gg\sigma_1$
below $T_c$ and at low frequency, in which case one obtains the relations
\begin{eqnarray}
\label{eq:local}
R_s(T) & = \frac{\mu_0^2}{2}\omega^2\lambda^3(T)\sigma_1(\omega,T) \\
X_s(T) & = \mu_0\omega\lambda(T) \nonumber \ \ \ \ \ \ .
\end{eqnarray}
A more detailed discussion of the electrodynamics and the
extraction of $\sigma_1(\omega,T)$ from $Z_s(T)$ will be presented in 
Section \ref{sec:results}, but the equations above are useful for quick
estimates and for understanding the main features of the microwave properties
of these superconductors. 

The expression for $R_s(T)$ in Eq. \ref{eq:local} embodies what is most
difficult about the microwave measurements. It contains both $\sigma_1$ and a
term of the form $\omega^2\lambda^3$, and physically can be interpreted as
the microwave
absorption processes ($\sigma_1$) occuring within the rather shallow depth 
into which the microwaves penetrate (hence the term $\omega^2\lambda^3$).
This screening length set by the superfluid makes $R_s$ extremely small
below $T_c$,  such that for small single crystals, one is forced to employ
cavity perturbation in very high
Q microwave resonators. This is most often achieved by the use of
microwave cavities made from conventional superconductors cooled to low
temperature. This fixed frequency type of measurement must then be performed
with several different resonators if one is to build up a reasonably complete
picture of the microwave conductivity spectrum $\sigma_1(\omega)$.

The data presented here involve measurements with 5 different resonators
spanning a range from 1 to 75 GHz and utilizing a number of variations on
the basic method of cavity perturbation. The common feature is that all of
the measurements have been performed on the same sample, a thin plate of
$YBa_2Cu_3O_{6.993}$ oriented such that the microwave magnetic field of each
cavity lay in the plane of the plate ($\vec{H}_{rf}\parallel\hat{b}$).
This geometry has the advantage that demagnetization
factors are quite small, so the surface current distributions are fairly
uniform and are similar in all of the measurements, making comparison from
frequency to frequency quite reliable. The disadvantage of this geometry is that
although it mainly measures the surface impedance for currents running across
the crystal face in the $\hat{a}$ direction, 
there is some admixture of the $\hat{c}$-axis surface impedance coming from 
currents running down the thin edge of the crystal. However, we have previously
shown that these effects are small for a thin crystal, because the temperature
dependences of $R_s(T)$ and $X_s(T)$ are quite weak in the $\hat{c}$-direction,
except near $T_c$\cite{bonn96,hosseini98}. The absolute value of $R_s(T)$ for
the $\hat{c}$-direction is
also relatively low \cite{hosseini98}.

The measurements at 1 GHz were performed in a loop-gap resonator initially
designed for measurement of $\lambda(T)$ \cite{hardy93}.
Like most of the resonators used
in this study, the loop-gap is plated with a Pb:Sn alloy which is
superconducting below 7 Kelvin and has very low microwave loss at 1.2 Kelvin.
In the case of the 1GHz loop-gap, the cavity Q can be as high as $4\times 10^6$
at low temperature.  Another common feature of all of the measurements is that
the sample is mounted  on a thin sapphire plate with a tiny amount of
silicone grease,
with the thermometry and sample heater located at the other end of
the sapphire plate, outside of, and thermally isolated from the resonator.
In this way the resonator can be held fixed at the regulated $^4He$ bath 
temperature while the sample temperature is varied. A unique feature of the 
1 GHz loop-gap system is that the sample is held fixed in the resonator, with
the sapphire plate and thermometry stage supported by a thin walled quartz tube
which sustains the temperature gradient between the $^4He$ bath  
and the sample. This means that the sample cannot be removed from the 
resonator during the measurements, but we find this restriction
is necessary  for the high precision
measurements of $\lambda(T)$ which rely on the sample being held rigidly in
a position that does not vary when the temperature is changed. Without this
type of construction,
motion of the sample and sapphire plate in the fields of the
resonator can give rise to experimental artifacts. This is
because $\lambda(T)$ is
determined from very small changes in the resonator frequency as the
sample temperature is changed. This technique gives highly precise and 
reliable temperature dependences relative to the base temperature, but
is limited to measuring differences;
$\Delta\lambda(T)=\lambda(T)-\lambda(1.2\,K)$ and 
$\Delta R_s(T)=R_s(T)-R_S(1.2\,K)$. In this paper we will not be focussing
on the low temperature limit $R_s(T\rightarrow 0)$, which is a very small
microwave loss in comparison to the loss associated with thermally 
excited quasiparticles. On the other hand, we do need to know  
$\lambda(1.2\,K)$ in order to extract microwave conductivities, so we
obtain this value from infrared measurements \cite{tallon}.

The 2 GHz measurements have been performed in a Nb split-ring resonator that
will be described in more detail elsewhere\cite{gowe98}.
Its geometry is related to the
1 GHz loop-gap resonator, with the key operational
difference being that the sample can be
pulled in and out of the resonator to measure the Q and resonant frequency 
of the unloaded cavity.
Measurements at 13.3, 22.7, and 75.1 GHz have been performed in the axial 
microwave magnetic fields of the $TE_{011}$ modes of three right-circular
cylindrical cavities. In these higher frequency cavity perturbation
measurements, the sample can also be
withdrawn from the resonator.
This allows one to determine $R_s(T)$ from
$R_s(T)\propto 1/Q_s-1/Q_0$ where $Q_0$ is the Q of the
empty cavity and $Q_s$ is the Q with the sample inserted. A small correction
for other sources of loss is made by measuring the sapphire holder and grease
without the sample; the calibration of the absolute value of the
surface resistance is obtained by measuring a Pb:Sn reference sample whose
normal state surface resistance is governed by the classical skin
effect.

The sample used for all of these measurements was a single crystal of
$YBa_2Cu_3O_{6.993}$ grown by a flux growth technique using $BaZrO_3$
crucibles\cite{liang98}.
The $BaZrO_3$, which does not corrode during the crystal
growth, yields crystals of much higher purity \cite{liang98,erb95}
(better than 99.99\%)
than those grown in
more commonly used crucibles such as yttria-stabilized zirconia and alumina. 
The sample used for this was detwinned at 250 C under uniaxial stress and was
subsequently annealed at (350 C) for 50 days to produce a sample with nearly
filled chain oxygen sites. This produces a sample with a slightly lower $T_c$
(88.7 K) than the maximum obtained near an oxygen concentration of
6.91 \cite{liang98},
but provides particularly defect-free samples without the
oxygen vacancy clustering discovered by Erb et al. \cite{erb96b}.
The dimensions of the
sample were initially 2x1x0.02 $mm^3$ for the 1 GHz measurements, which have
been reported elsewhere \cite{kamal98}.
Subsequent measurements in the higher frequency
resonators were performed on smaller pieces cleaved from the original
crystal.

\section{Experimental Results and Analysis}
\label{sec:results}

\begin{figure}[t]
\begin{center}
\leavevmode
\epsfxsize=3in %% 
\epsfbox{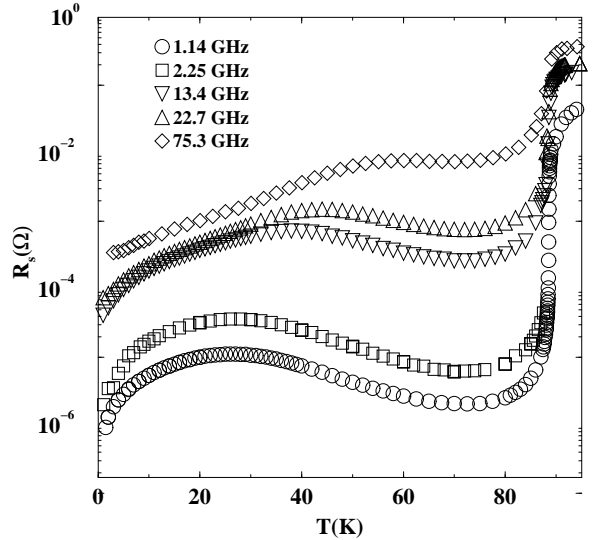}
\end{center}
\caption{The surface resistance $R_s(T)$ for currents running in the
$\hat{a}$ direction of a high purity crystal of $YBa_2Cu_3O_{6.993}$ 
The measurements range over 5 different frequencies from 1 to 75 GHz.}
\label{fig:rs1}
\end{figure}

\begin{figure}[t]
\begin{center}
\leavevmode
\epsfxsize=3in %% 
\epsfbox{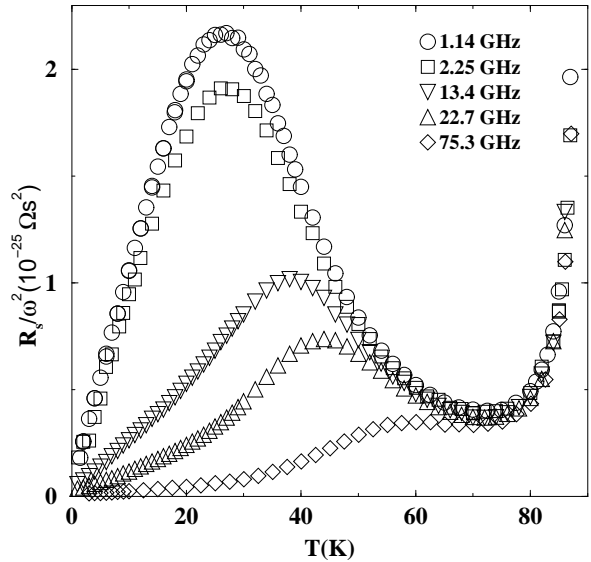}
\end{center}
\caption{The same measurements of surface resistance as shown in
Fig. 1, but with the behaviour below $T_c$ emphasized by dividing out
a frequency-squared dependence associated with superfluid screening.}
\label{fig:rs2}
\end{figure}

Fig. \ref{fig:rs1}
shows the microwave surface resistance of 
$YBa_2Cu_3O_{6.993}$ for currents running in the 
$\hat{a}$ direction. At all 5 of the frequencies
shown in the figure, the rapid drop in $R_s(T)$ below $T_c$ is due to the
onset of screening by the superfluid, with the overall magnitude of the drop 
depending strongly on frequency as expected from the term in $R_s(T)$ that 
varies as $\omega^2\lambda^3(T)$. 
For easier comparison of the
different frequencies, it is convenient to plot the low loss data in the
superconducting state as $R_s/\omega^2$ versus temperature, as shown in 
Fig. \ref{fig:rs2}. 
One feature of these figures is that they demonstrate
that above 65 K the loss scales quite closely as $\omega^2$, indicating
that $\sigma_1$ is frequency independent above 65 K in the microwave range.
Alternatively, if one expects $\omega^2$ scaling, then this indicates
that the comparison between the frequencies is working extremely well,
even though the data is generated in 5 different resonators with separate
calibrations of the absolute value in each experiment.
The curves in Fig. \ref{fig:rs2} show a number of qualitative features
that have already been observed in earlier microwave measurements of this
material. $R_s(T)$ exhibits a broad peak which shifts up in temperature and 
diminishes in size as the measurement frequency is increased. Also noteworthy
is that we continue to observe {\it nearly} linear behaviour of $R_s(T)$ at low
temperature and low frequency. 

The broad peak in $R_s(T)$ was originally attributed to a quasiparticle
scattering time $\tau$
that increases rapidly with decreasing temperature 
below $T_c$ and
competes with a density of these thermally excited qusiparticles that
decreases with temperature \cite{bonn92}.
At intermediate temperatures from $T_c$
down to about 30 K
the quasiparticle lifetime increases much
more rapidly than the density decreases, causing $\sigma_1(T)$
(and thus $R_s(T)$)
to rise with decreasing temperature. For the early measurements near 2 GHz it
was suggested that the quasiparticle scattering time reaches a limiting value
near 30 K, possibly due to impurity scattering, at which point the 
quasiparticle density takes over and causes $R_s(T)$ to fall again with
further decreases in temperature \cite{bonn93}.
The speculation that impurities are involved
in the turnover at 30 K was partly checked by studying samples doped with
Ni and Zn \cite{bonn94}. 
These doping studies showed either a smaller peak shifted to
higher temperature or no peak at all, consistent with the quasiparticle
scattering rate running into an impurity limit at higher temperature, even when
only 0.15\% Zn was added to the sample. This sensitivity
to such low levels of impurities raises a serious concern
over these earlier crystals grown
in yttria-stabilized zirconia crucibles, because during crystal growth
the residual impurity level
due to uptake of material from the corroding crucible reaches the
0.1\% level. The results shown here on a new higher purity sample confirm
the original speculation that it is this residual impurity scattering that
limits the increase in quasiparticle 
lifetime, even in quite pure crystals. The low frequency
surface resistance of the new
crystal rises considerably higher above the minimum near
70 K than was observed in earlier measurements of samples at 2 and 4 GHZ
and reaches its maximum at a lower temperature \cite{bonn93}.
Both of these quantitative changes are consistent with the new samples
simply having lower impurity scattering and so reaching a higher 
quasiparticle lifetime limit at a lower temperature than was seen in the 
yttria-stabilized zirconia grown crystals.

The frequency dependence of the peak in $R_s(T)$ provides further evidence
of the rapid increase in the quasiparticle lifetime. In much higher
frequency measurements on thin films, Nuss et al. \cite{nuss91}
were the first to observe a similar broad
peak in the THz range that shifted up in temperature and decreased in size
at higher frequencies. They pointed out that this could be accounted for by 
relaxation effects. If the quasiparticle lifetime increases sufficiently that
$\omega\tau\ge 1$, then the sample enters a relaxation
regime where $\sigma_1(T)$ falls
as $\tau$ continues to increase. So for high frequency measurements the 
temperature at which $R_s(T)$ reaches its peak roughly indicates where
$\omega\tau =1$ at each measurement frequency. The fact that we already observe
these relaxation effects at 13 GHz indicates that the scattering time $\tau$
grows to extremely large values in the new high purity samples.

\begin{figure}[t]
\begin{center}
\leavevmode
\epsfxsize=3in %% 
\epsfbox{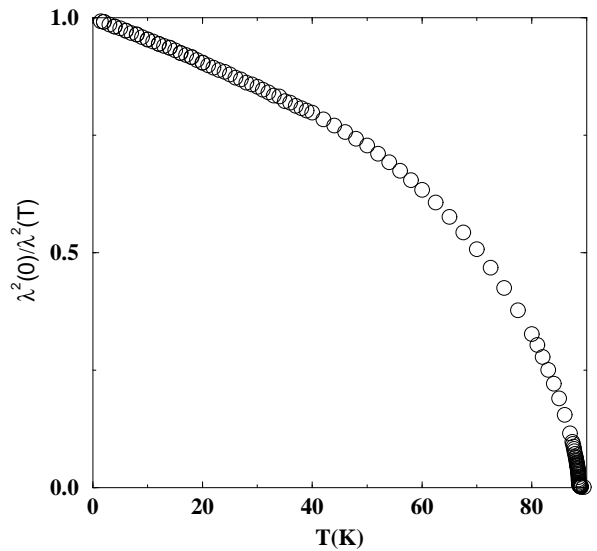}
\end{center}
\caption{The temperature dependence of the $\hat{a}$ axis penetration depth
 of $YBa_2Cu_3O_{6.95}$
 plotted as $1/\lambda^2(T)$, which is a measure of the superfluid
 density via $n_se^2/m^*=1/(\mu_0\lambda^2)$.}
\label{fig:lambda}
\end{figure}

To better understand the data it is desireable to extract $\sigma_1(\omega,T)$
from these measurements of $R_s(\omega,T)$. At the lowest frequency of
1 GHz, where
$\omega\tau\ll 1$ and we have simultaneous measurements of 
$R_s(T)$ and $\lambda(T)$,
it is straightforward to extract 
$\sigma_1(T)$ with only an assumption that the electrodynamics are local.
The 1 GHz measurements of $\lambda(T)$ used in this analysis
are shown in Fig. \ref{fig:lambda},
plotted as $1/\lambda^2(T)$.
The screening by the superfluid follows the local London model if 
a superconductor is in the limit $\xi\ll\lambda$. Strictly
speaking, this local limit is a bit more complicated
for the case of a superconductor
with nodes in the energy gap because the coherence length is then $k$-dependent
and becomes large in the node directions. However, the consequences of this
type of non-locality for the electrodynamics would only be observable at low
temperatures and
the effect would be small in the measurement geometry used 
here \cite{kosztin97}.
A much more serious problem for the data analysis is that at the higher 
frequencies where $\omega\tau\geq 1 $, the thermally excited quasiparticles
also contribute to the screening of microwave fields and it is then incorrect
to use the 1 GHz measurements of $\lambda(T)$ in order to extract $\sigma_1(T)$ 
from $R_s(T)$. Effectively, the penetration depth becomes frequency
dependent, a phenomenon that has been observed directly in mm-wave
measurements on particularly high quality thin films \cite{dahne95}.
Ideally this problem can be solved by measuring both $R_s(T)$ and 
$X_s(T)$ at each measurement frequency, but for our higher frequency resonators
there are considerable technical difficulties with this measurement. 

It is possible to work around the fact that we only have measurements of
the real and imaginary part at one of the frequencies, as long as there
is adequate information on the frequency dependence of the real part of
the surface impedance. The problem is analogous to the one commonly
faced in far infrared spectroscopy of opaque samples,
where the reflectance, but not the phase of the reflected light, is
measured over a wide frequency range. Kramers-Kronig relations are the
usual solution if only one of the
optical constants is known, but is known over a wide frequency range.
Such relations are a consequence of causality in linear response theory
and connect one optical constant (such as reflectance)
to its partner (phase of reflected light), via an integral over all frequecies.
In principle, data at the five microwave frequencies presented here could
be connected to far infrared measurements of reflectance in order to 
perform this analysis and extract $\sigma_1(\omega)$ and $\sigma_2(\omega)$.
However, such an analysis is somewhat difficult because
there still exists a substantial gap in the mm-wave frequency range
between our highest frequency measurement and the lowest frequency far
infrared measurements on crystals. Some data has been obtained in the
mm-wave region using techniques such as time domain THz measurements 
\cite{nuss91,spielman94,pimenov98} and
direct infrared absorption \cite{miller93},
but these measurements have all been performed 
on films rather than untwinned single crystals. The quasiparticle scattering
rate is typically much higher in films than it is in single crystals, so data
taken on such different samples cannot be analyzed together in this way.
An alternative to a Kramers-Kronig analysis is to fit the frequency dependent
surface resistance to a model, but this is not a very satisfactory
procedure if one only has data at five microwave frequencies and
one doesn't know the shape of the conductivity spectrum
{\it a priori}.  A further problem with both of these techniques is that
the far infrared data is only available at a couple of temperatures below
$T_c$, so they can not be used to do a complete analysis of the data
presented here.

Our main approach to analyzing the surface resistance data is to use
the $R_s(T)$ measurements at 1 GHz to arrive at an estimate of how
much screening by the thermally excited quasiparticles must be included
when extracting $\sigma_1(T)$ from $R_s(T)$ at higher frequencies. 
Although the procedure involves some assumptions about the model for the
screening, we will show that the corrections are small enough that
the effect of uncertainty in the choice of model does
not significantly 
affect the conductivities that we extract in the analysis. We begin by
writing down a general 2-fluid expression for the microwave conductivity that
includes contributions to the real and imaginary part from both the
superfluid and normal fluid (the conductivity due mainly
to thermally excited
quasiparticles),
\begin{eqnarray}
\sigma(\omega,T)=\sigma_{1S}-i\sigma_{2S}+\sigma_{1N}-i\sigma_{2N} \\
\ \ \ \ \    = \frac{n_se^2}{m^*}\delta(\omega) - i\frac{n_se^2}{m^*\omega} +
   \sigma_{1N}-i\sigma_{2N} \nonumber
\label{eqn:twoflu}
\end{eqnarray}
where $\sigma_{1S}$ and $\sigma_{2S}$ are the real and imaginary parts of the
superfluid conductivity and $\sigma_{1N}$ and $\sigma_{2N}$ are the real and 
imaginary parts of the normal fluid conductivity. The superfluid contribution
consists of a delta-function at $\omega = 0$ with an oscillator strangth given
by the superfluid density divided by the effective mass ($n_s/m^*$),
and a Kramer-Kronig related imaginary part that is
the inductive response of the superfluid at nonzero frequencies. This
superfluid response can be expressed in terms of the penetration depth using
$n_se^2/m^*=(\mu_0\lambda^2(T))^{-1}$, and away from $\omega=0$ the 
delta-function can be neglected, leaving
\begin{equation}
\sigma(\omega,T)= \sigma_{1N}(\omega,T) -i\left[ \sigma_{2N}(\omega,T)
+\frac{1}{\mu_0\omega\lambda^2(T)}\right] .
\label{eqn:twoflu2}
\end{equation}
Thus, in general the real part of the conductivity comes from the normal fluid
response and the imaginary part has contributions from both the normal and
superfluid, although the superfluid dominates the imaginary part at low 
frequency.

In the limit of local electrodynamics the connection between
this conductivity and the surface impedance is made via
\begin{equation}
Z_s=R_s+iX_s=\left(\frac{i\mu_0\omega}{\sigma_1-i\sigma_2}\right)^{\frac{1}{2}}.
\label{eqn:zs}
\end{equation}
Thus, at 1 GHz where we have simultaneous measurements of both $R_s$ and $X_s$,
$\sigma_1$ and $\sigma_2$ can be extracted using
\begin{eqnarray}
\sigma_1=2\mu_0\omega\frac{R_sX_s}{(R_s^2+X_s^2)^2} \\
\sigma_2=\mu_0\omega\frac{X_s^2-R_s^2}{(R_s^2+X_s^2)^2} \nonumber .
\label{eqn:sigma1}
\end{eqnarray}

At higher frequencies, where we only have measurements of $R_s(T)$, it is
useful to write $\sigma_1$ in terms of $R_s$ and $\sigma_2$ in the
following way \cite{bonn93}:
\begin{eqnarray}
\label{eqn:sigma2}
\sigma_1=\left[\left[\frac{\sigma_s}{2}\pm\left(\frac{\sigma_s^2}{4}-
\sigma_2\sigma_s\right)^{1/2}\right]^2 -\sigma_2^2\right]^{1/2} \\
{\rm with\ the\ sign\ choice\ + (-)\ for}\ \sigma_1 > (<) \sqrt{3}\sigma_2
\nonumber \\
{\rm and\ where}\ \sigma_s = \frac{\mu_0\omega}{2R_s^2} . \nonumber
\end{eqnarray}
In the normal state, where $\sigma_1 \gg \sigma_2$ at low frequencies,
this expression reduces to the classical skin effect result 
$\sigma_1=\mu_0\omega/2R_s^2$. The expression in Eq. \ref{eqn:sigma2}
continues to be valid right through the superconducting transition and can
be used to extract $\sigma_1$ from measurements of $R_s$ provided that one
also has some information on $\sigma_2$. The extent to which one can use this
expression in the superconducting state
depends upon the relative importance of the normal fluid contribution
to $\sigma_2$. In regimes where this contribution is small, $\sigma_2$ is
simply given by the superfluid response 
$\sigma_{2S}=(\mu_0\omega\lambda^2(T))^{-1}$ which can be calculated from the
penetration depth measurements. To clearly illustrate the problem that occurs
when the normal fluid contribution to $\sigma_2$ is not small, we examine a 
particular version of the 2-fluid model where the normal fluid conductivity
is assumed to have a Drude frequency dependence:
\begin{equation}
\sigma_{1N}-i\sigma_{2N} = \frac{n_ne^2}{m^*}\left[\frac{\tau}{1+(\omega\tau)^2}
-i\frac{\omega\tau^2}{1+(\omega\tau)^2}\right]
\label{eqn:drude}
\end{equation}
where $\tau$ is the scattering time of the normal fluid and $n_n/m^*$ is the
normal fluid density over the effective mass. The ratio of the normal fluid
and superfluid contributions to the effective screening is then
\begin{equation}
\frac{\sigma_{2N}}{\sigma_{2S}} = 
\frac{n_n}{n_s}\frac{(\omega\tau)^2}{1+(\omega\tau)^2} .
\end{equation}
The normal fluid contribution to screening is thus unimportant at low frequency
($\omega\tau\ll 1$) or when the normal fluid density is small ($n_n/n_s\ll 1$,
which occurs at low temperatures). Conversely, difficulties arise when 
$\omega\tau > 1$ and $n_n/n_s$ is not small, which is likely the case for our
data at 13 and 22 GHz in the temperature range from 20--40 K, and for the 
75 GHz data from 20--60 K. 

For the conductivities that we will show below we take the following 
approach to estimating the normal fluid contribution to $\sigma_2$. At 1 GHz
$\sigma_1$ is calculated directly from the simultaneous measurements of
$R_s(T)$ and $X_s(T)$. Then, in the spirit of our earlier analysis of this type
of data, we extract the temperature dependent normal fluid scattering time 
$\tau(T)$ using the Drude model above (Eq. \ref{eqn:drude}). The normal fluid
density is calculated from the penetration depth measurements by assuming 
that 
\begin{equation}
\frac{n_ne^2}{m^*}\left( T \right) +\frac{n_se^2}{m^*}\left( T \right)
= \frac{n_se^2}{m^*} \left( T=0 \right) .
\label{eqn:normalfluid}
\end{equation}
which amounts to assuming that all of the low
frequency oscillator strength is being shifted from the normal fluid to the
superfluid as the temperature is decreased. The $\tau(T)$ and $n_n(T)$ derived
from the 1 GHz data can then be used to make an estimate of the normal fluid
contribution to $\sigma_2$ needed to analyze the data at higher frequencies.
When this analysis is performed, we find that the normal fluid screening
influences
the extraction of $\sigma_1$ from $R_s(T)$ only 
at the level of 20\% or less. This
can be understood if one notices in the $R_s(T)$ that $\omega\tau\sim 1$ at
temperatures of 50 K or less at which point the normal fluid density has already
fallen below 20\% of the total oscillator strength. Thus, we find that the
effect of the normal fluid screening is actually fairly small in the microwave
range, so any uncertainty associated with the choice of a Drude model to 
account for this screening does not have a serious impact on the values of
$\sigma_1$ that we derive from the data at 13, 22, and 75 GHz. Furthermore,
the values of $\tau(T)$ that we use in the analysis above turn out to be
consistent with the conductivity spectra $\sigma_1(\omega)$ that will be
discussed below, so the correction for screening by the normal fluid is
self-consistent.

\begin{figure}
\begin{center}
\leavevmode
\epsfxsize=3in %% 
\epsfbox{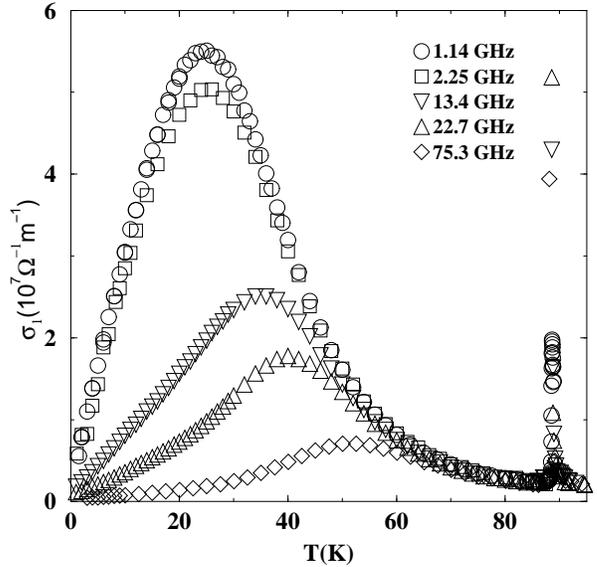}
\end{center}
\caption{The temperature dependence of the $\hat{a}$-axis
microwave conductivity of $YBa_2Cu_3O_{6.993}$ extracted from the 
surface resistance measurements of Fig. 1. The sharp spike near $T_c$
is a result of superconducting fluctuations and the broad peak at lower
temperatures is caused by the increase in the scattering time of thermally 
excited quasiparticles.}
\label{fig:sigmaT}
\end{figure}

Fig. \ref{fig:sigmaT} shows the conductivities extracted from the $R_s(T)$
data of Fig. \ref{fig:rs1}, using the methods described above. 
The sharp
upturn at $T_c$ marks the presence of superconducting fluctuations, which
have been discussed in greater detail by Kamal et al.\cite{kamal94}
and Anlage et al. \cite{anlage95} and
are not the main focus of the work presented here. A number of previous 
measurements on earlier generations of $YBa_2Cu_3O_{7-\delta}$ crystals
have shown the presence of
an extra
sample-dependent peak in $\sigma_1(T)$ just below $T_c$, a feature that
was discussed by Olson and Koch \cite{olson91} and by
Glass and Hall \cite{glass91}
who attributed it to having a sample with a 
broadened transition. A similar feature was also reported recently 
by Srikanth et al. for $BaZrO_3$-grown crystals \cite{srikanth97},
but since we see no sign of this in our new high purity
samples, we conclude that such features are associated with a spread in $T_c's$
in the surface of the sample. 
The main feature that we do observe in the conductivity is that
$\sigma_1(T)$ has a large broad peak, 
rising to nearly 25 times the normal state conductivity. 
The conductivity rises a factor of two higher than was seen in measurements
at 2 and 4 GHz on an earlier generation of crystals grown in yttria-stabilized
zirconia crucibles \cite{bonn94}.
The peak also appears at a lower temperature, 25 K instead
of the 35 K turnover observed in the lower purity crystals. This effect
of very low levels of impurities is not consistent with the early suggestion of
Klein et al. \cite{klein94}
that this feature is the result of BCS-type coherence effects. A
coherence peak, essentially a density of states effect, is observed near $T_c$
in $\sigma_1(T)$ of s-wave superconductors such as Pb\cite{holczer91}. However,
the peak observed here at 1 GHz is much too large and too low in temperature to
be attributed to such an effect. Furthermore, a strong coherence peak has not
been seen in NMR measurements of $T_1$ in this material \cite{hammel89}
nor is it expected in 
a d-wave superconductor, the now widely accepted pairing state of 
$YBa_2Cu_3O_{7-\delta}$. \cite{scalapino90,bulut92} 

Thus, in the absence of strong coherence effects, 
we have attributed the rise in 
$\sigma_1(T)$ below $T_c$ to a rapid increase in the scattering
time $\tau$ of thermally excited quasiparticles, as discussed in the
Introduction.
The fact that this peak rises higher and turns over at a lower
temperature in the new, higher purity crystals is consistent with this
interpretation. That is, in the higher purity sample, $\tau$ runs into its
impurity limit at a somewhat lower temperature than it did in the earlier 
generations of crystals and this impurity-limited scattering 
time is very large in
the new $BaZrO_3$-grown crystals. An estimate of the increase can be made as 
follows.
The penetration depth measurements indicate that more than
90\% of the
normal fluid density is gone at 25 K (see Fig. \ref{fig:lambda}), so the 
25-fold increase in $\sigma_1$ between $T_c$ and 30 K implies an increase in
$\tau$ by at least a factor of 250 over the scattering time just above
$T_c$. This is actually an underestimate because not all of the far
infrared oscillator
strength in the normal state ends up condensed into the superfluid at low
temperatures. Such a huge increase in lifetime is consistent with the relaxation
effects observed in the data at 13 GHz and higher. As a rough illustration of
this agreement, far infrared measurements indicate that $\omega\tau\sim 1$
at about 3000 GHz just above $T_c$, so a 300-fold increase in $\tau$ below
$T_c$ would mean $\omega\tau\sim 1$ at about 10 GHz, leading to the
considerable
frequency dependence in $\sigma_1(\omega)$ that we observe
in the microwave range.
\begin{figure}[t]
\begin{center}
\leavevmode
\epsfxsize=3in %% 
\epsfbox{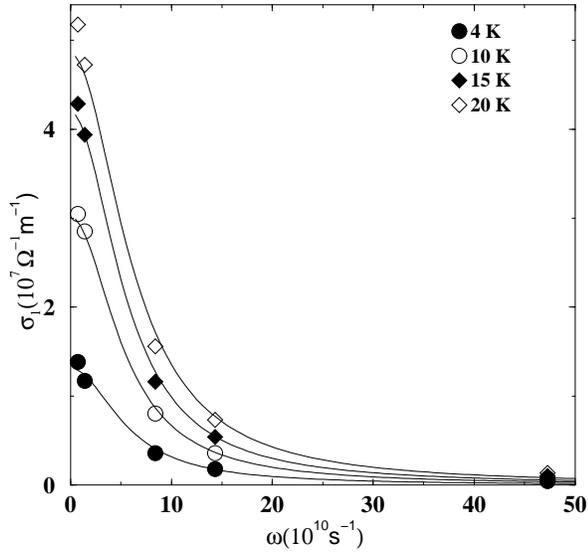}
\end{center}
\caption{The conductivity spectrum at 4 selected temperatures between
4 and 20 K,
extracted from the $\sigma_1(T)$ curves of Fig. 4. In this 
temperature regime, the conductivity due to thermally excited quasiparticles
has a nearly temperature independent width of 8 GHz and a nearly temperature
independent shape that is close to a Drude lineshape (the lines are Drude
fits).}
\label{fig:sigmaw1}
\end{figure}

\section{Conductivity Spectra and Quasiparticle Lifetime}
\label{sec:tau}
\begin{figure}[t]
\begin{center}
\leavevmode
\epsfxsize=3in %% 
\epsfbox{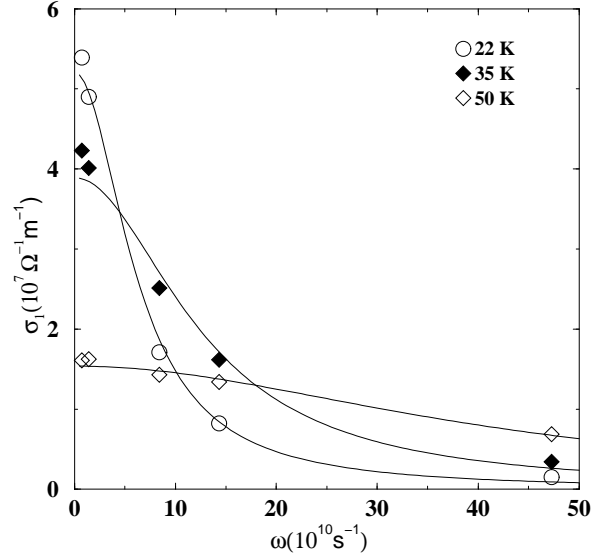}
\end{center}
\caption{The conductivity spectrum at 3 selected temperatures above
20 K,
extracted from the $\sigma_1(T)$ curves of Fig. 4. Above 20 K, the width of
the conductivity peak broadens rapidly, stretching out of the microwave
frequency range above 55 K. 
These conductivity spectra continue to be reasonably well fit by Drude 
lineshapes, as shown by the lines in the figure.}
\label{fig:sigmaw2}
\end{figure}
The evolution of the conductivity with temperature is better illustrated in
Figs. \ref{fig:sigmaw1} and \ref{fig:sigmaw2}
where we show the conductivity spectrum $\sigma_1(\omega)$
at several different temperatures. 
In fact, the central technical achievement
of this work is that we now have measurements at enough frequencies that both 
the shape of $\sigma_1(\omega)$ and its temperature dependence are quite clear.
Fig. \ref{fig:sigmaw1} shows the conductivity spectrum at four different
temperatures below 25 K, the temperature range where we have argued above that 
the conductivity is dominated by thermally excited quasiparticles that are
being scattered by a low level of impurities. We will not concentrate here on 
the lowest temperature data, where the loss becomes very small and difficult to 
measure accurately by most cavity perturbation techniques. 
Fig. \ref{fig:sigmaw1} shows that from 4 to 20 K, the conductivity consists
of a very narrow peak whose width is about 8 GHz. The lines in the figure are
fits to a Drude lineshape (the real part of Eq. \ref{eqn:drude}),
demonstrating that the
conductivity spectrum of the thermally excited quasiparticles is Drude-like.
There are deviations from this lineshape; in particular, the data at the
lowest frequencies tends to lie above any Drude curve that manages to
fit the data at 13 GHz and above. That is, $\sigma_1(\omega)$ seems
slightly more cusp-like than a Lorentzian curve.
Despite these somewhat subtle details, the nearly Lorentzian lineshape
largely confirms the ansatz originally used by Bonn et al. \cite{bonn93}
to analyze the early
$R_s(T)$ measurements. Now, however, it is possible to glean 
an effective scattering 
rate $1/\tau$ by fitting $\sigma_1(\omega)$ to a Drude lineshape, rather than 
assuming one. The fits shown in Fig. \ref{fig:sigmaw1} all yield a linewidth
close to 8 GHz, so
the dominant change in the spectra in this temperature range seems to be an 
increase in oscillator strength, due to the
shift of spectral weight from the superfluid response (a
$\delta$-function at $\omega$=0) to the microwave conductivity.
\begin{figure}
\begin{center}
\leavevmode
\epsfxsize=3in %% 
\epsfbox{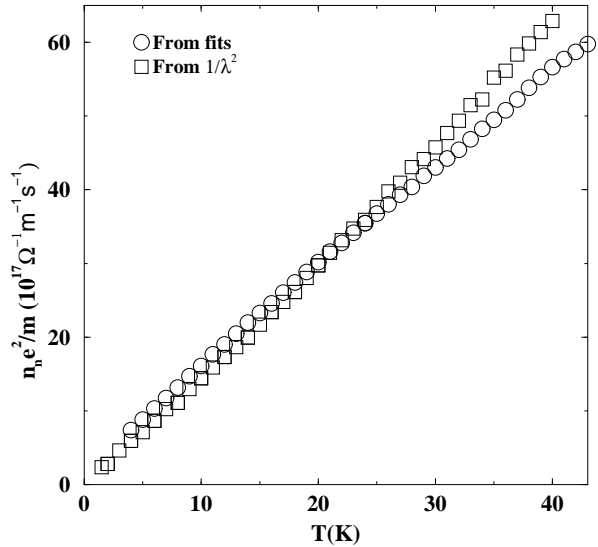}
\end{center}
\caption{A comparison of the normal fluid oscillator strength determined
in two ways; from Drude fits to spectra like those of Figs. 5 and 6
and from the disappearance of oscillator strength
in the superfluid response, as measured by $1/\lambda^2$ (Fig. 3).}
\label{fig:normalfluid}
\end{figure}

One sees in Fig. \ref{fig:sigmaw2} that above 25 K the conductivity peak starts
to broaden rapidly and by 60 K the width becomes much greater than the
frequency range of the microwave measurements, so we have no direct measure
of the width and shape of $\sigma_1(\omega)$ above 60 K.  
Again, in this figure the lines are fits to a Drude lineshape. The fits are
not perfect, but the flattening of the curve at the lowest two frequencies and 
the rapid roll-off with increasing frequency are key features of a Drude 
conductivity spectrum.
Although the Drude lineshape is not a perfect fit to the data (even with the 
5-10\% uncertainty in the absolute value at the various frequencies),
overall the fits between 4 and 55 K are good enough 
to consider plotting the fit parameters as a function of temperature. 

An interesting cross-check of these fits to $\sigma_1(\omega)$ is to compare
the oscillator strength in the normal fluid conductivity peak, which is 
given by the fit parameter $n_ne^2/m^*$, to the superfluid density 
$n_se^2/m^* = (\mu_o\lambda^2)^{-1}$
extracted from the penetration depth measurements. If one assumes
that all of the oscillator strength ends up in the superfluid $\delta$-funtion
as T$\rightarrow$0, then the superfluid density is related to the normal
fluid density via Equation \ref{eqn:normalfluid}. 
Fig. \ref{fig:normalfluid} shows a comparison between the normal fluid
density inferred from the fits to the $\sigma_1(\omega)$ peaks and the
normal fluid density inferred from the 
penetration depth via Eq. \ref{eqn:normalfluid}.
\begin{figure}
\begin{center}
\leavevmode
\epsfxsize=3in %% 
\epsfbox{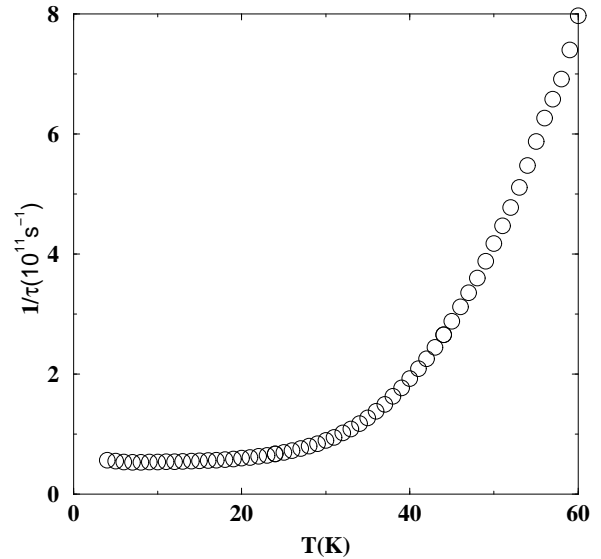}
\end{center}
\caption{The scattering rate of the thermally excited quasiparticles,
as inferred from the width of Drude fits to the conductivity
spectra of Figs. 5 and 6.}
\label{fig:tau1}
\end{figure}
This figure indicates that the normal fluid density does track the 
decline of the superfluid density with increasing temperature, which lends
some degree of confidence to this parameterization of the data. The agreement
at low temperatures is excellent, with an increasingly serious
deviation betweeen the
two above 30 K. The lack of agreement at the higher temperatures is an 
indication that 
the Drude lineshapes do not completely keep track of where all of the oscillator
strength is going as temperature increases. 
There is some deviation at low temperatures, taking the form of a 
normal fluid oscillator strength
that is extrapolating linearly
towards a non-zero value as T$\rightarrow$0. This is an indication of
the presence of residual conductivity in the low temperature, low frequency
limit, which is expected for a d-wave superconductor 
\cite{lee93}. 

Although the comparison of the oscillator strengths
shown in Fig. \ref{fig:normalfluid}
indicates that the Drude fits to $\sigma_1(\omega)$ do not perfectly account
for the redistribution of oscillator strength, the shape does provide a 
reasonable measure of the width of the peaks
from 4 to 55 K. The temperature dependent width coming
from these fits, which we have previously interpreted
as the scattering rate of the thermally excited quasiparticles, is shown in 
Fig. \ref{fig:tau1}. 
One of the key results of these measurements is that
the width of the normal fluid peak is very small, only about 8 GHz,
and it is nearly temperature-independent up to
20 K. The 8 GHz width suggests a very long
quasiparticle 
scattering time of 2$\times 10^{-11}$ s.
In previous work performed on lower purity samples
measured at only 2 frequencies (4 and 35 GHz) a similar
conclusion was drawn regarding the spectrum of the thermally excited 
quasiparticles. In that 
data it was suggested that the width was about 30 GHz,
with little temperature dependence of the 
scattering rate up to somewhat higher
temperature, about 30 K. The narrower width of 8 GHz lasting only up to 
about 20 K is consistent with the higher purity of the new $BaZrO_3$-grown
crystals. That is, the lower impurity limit to the width makes the broadening
due to inelastic scattering apparent at somewhat lower temperatures.

Above 20 K the scattering rate increases very rapidly, broadening the peak in
$\sigma_1(\omega)$ out of the microwave range by 60 K. Eventually, this 
width will broaden out into the far infrared conductivity spectrum
in the normal state, a width of $\sim 2k_BT$ which is about 3600
GHz just above $T_c$. That is, the quasiparticle scattering rate falls
by a factor of about 500 from $T_c$ down to the impurity limit encountered
at 20 K. Thus, the exceptionally low level of defects in the new $BaZrO_3$
grown crystals, coupled with the measurements at 5 microwave frequencies,
allows us to determine the temperature dependence of the inelastic scattering
rate over a wide temperature range. Qualitatively similar results have been
obtained in measurements on thin films by high frequency microwave and THz
techniques \cite{nuss91,pimenov98}.
However, the much higher level of defects in such samples limits
the temperature range over which the evolution of the inelatic scattering 
rate can be observed. In the high purity crystals, the temperature
dependent scattering can be tracked all the way down to 20 K.

\section{Discussion}
\label{sec:discussion}

For the purposes of discussing the data, we have divided the conductivity
spectra up into the two regimes discussed in the previous section. In the 
range below about 20 K where the width of the peak in $\sigma_1(\omega)$
is narrow and nearly temperature independent, studies of samples over a
wide range of purities indicate that this regime is governed by thermally 
excited quasiparticles being scattered by impurities or other defects
\cite{bonn94,zhang93}.
Waldram et al. have pointed out the possibility that non-local effects
might come into play in the conductivity in this regime, leading to an 
effective scattering rate that is not influenced by the density of residual
impurities \cite{waldram97}.
However, the considerable narrowing of $\sigma_1(\omega)$ that
we have observed upon going from YSZ-grown crystals to the higher purity
$BaZrO_3$-grown crystals indicates that the samples are still in a regime
where impurities play a role in the low frequency
scattering. 
We have previously suggested an intuitively
appealing way to interpret the spectra in this regime, based on a specific
version of the two 
fluid model \cite{bonn93}. In this phenomenological picture,
the quasiparticles excited near the nodes in the energy gap have a
temperature-independent scattering rate due to elastic scattering by
impurities and a conductivity spectrum with a Drude lineshape whose width is
set by this scattering rate $1/\tau_i$
just like impurity scattering in a normal metal.
In the high purity samples this $1/\tau_i$ would 
correspond to a strikingly long
mean free path of 4 $\mu m$ if one takes the Fermi velocity to be
$v_F = 2\times10^7 cm/s$. 
In this particular two fluid model, the only source of temperature dependence
in the low temperature microwave conductivity is the density of the
thermally excited quasiparticles, which increases linearly with temperature.
This is a straightforward consequence of the
linear dispersion of the gap function near the nodes and is also intimately
connected to the linear temperature dependence of the penetration depth 
(they are Kramers-Kronig related). 

Parameterizing
the normal fluid response in this way, with a temperature dependent density 
and a scattering rate, has been partly justified by calculations of the
microwave conductivity of a $d_{x^2-y^2}$ superconductor \cite{hirschfeld93}.
However, it is not so obvious that a temperature independent scattering
rate is expected for impurity scattering of these thermally excited
quasiparticles near the nodes, since they obviously
differ greatly from free carriers at an ungapped
Fermi surface. In particular, it has been pointed out by Hirschfeld et al.
that elastic impurity scattering in this situation
should lead to a frequency and temperature dependent scattering rate because
of the restricted phase space into which the quasiparticles at the nodes can
scatter \cite{hirschfeld93,hirschfeld94}. So, with this possible 
conflict between theory and the phenomenological model in
mind, we will make some more detailed comparison between
the microwave conductivity data and the relevant theoretical calculations. 

The transport properties (both microwave conductivity and thermal
conductivity) of a $d_{x^2-y^2}$ superconductor have been the
subject of considerable theoretical effort recently 
\cite{lee93,hirschfeld93,hirschfeld93b,hirschfeld94,xu95,hensen97}.
This work builds on earlier calculations
of the transport properties of anisotropic superconductors, aimed primarily
at explaining and predicting the properties of heavy fermion superconductors
\cite{hirschfeld89,klemm88}. The high temperature superconductors now offer
an opportunity to test these calculations in a situation where we seem to have 
identified a relatively simple anisotropic pairing state. Such comparisons are
somewhat complex because the question of transport properties at low
temperatures is inherently
a question of understanding impurity effects. This is especially the case for
anisotropic superconductors because the presence of impurities has a strong
impact on the excitation spectrum near gap nodes, particularly in the limit of 
unitary scattering. 

A key effect of impurities in an anisotropic
superconductor is 
to produce a band of impurity states with some width $\gamma$, thus
giving the superconductor a non-zero density of states at the Fermi level.
One surprising consequence of these states is a universal conductivity 
limit at
low frequency as $T\rightarrow 0$, first pointed out by P.A. Lee \cite{lee93}.
This residual conductivity is independent
of impurity concentration, the result of a cancellation between the density
of states induced by the presence of the impurities and the 
transport lifetime associated with those states. A version of this universal
limit has been observed by Taillefer et al. in thermal conductivity measurements
of pure and Zn-doped $YBa_2Cu_3O_{6.9}$ below 1 Kelvin \cite{taillefer97}.
To the best of our knowledge, this limit has not yet been definitively observed
in microwave conductivity measurements, in part due to sensitivity problems in 
the type of cavity perturbation measurement being discussed in this article.
Instead, our 
main concern here will be the behaviour of this conductivity as
temperature and frequency are increased, which involves conductivities that are
substantially larger than the T$\rightarrow$0 limit and are thus easily 
measurable with the methods discussed above.
This microwave conductivity has been the subject of considerable theoretical
effort, including both numerical work and a number of analytical
results in certain 
limits \cite{hirschfeld93,hirschfeld93b,hirschfeld94,xu95,hensen97}.

One particularly well
studied limit involves the electrodynamic properties
at temperatures and frequencies below the
impurity bandwidth $\gamma$, in the unitary scattering limit (scattering
phase shift $\rightarrow \pi/2$). In this limit the impurity bandwidth is given 
roughly by $\gamma \sim (\Gamma \Delta )^{1/2}$ where $\Delta$ is the 
magnitude of the gap and $\Gamma$ is the elastic scattering rate that the
impurities would contribute to the normal state resistivity.
For $\omega , T < \gamma$, where the transport properties are dominated by
this impurity band, it has been shown that
both $\sigma_1$ and $\lambda$ vary as $T^2$ \cite{hirschfeld93,hirschfeld93b}.
This quadratic behaviour has been seen in Zn-doped samples of 
$YBa_2Cu_3O_{6.95}$, where it was found that
at a Zn impurity concentration as low as 0.15 \%
the crossover energy scale is already $\gamma >$ 4 K \cite{achkir93,bonn94}.
However, Zn substitution for planar Cu's is the only impurity that we have
found that clearly gives this unitary scattering behaviour. Ni 
substitution for Cu, Ca substitution for Y, and the chain oxygen vacancies
all seem to have much weaker effects, even at defect levels of 1\% or more
\cite{bonn96}.
The previous generation of YSZ-grown crystals showed only slight
curvature in $\lambda(T)$ and $\sigma_1(T)$ below
4 Kelvin and the new $BaZrO_3$-grown crystals show no hint of $T^2$
temperature dependence down to 1.2 Kelvin. The relative rarity of this
quadratic behaviour (though it is common in thin films for reasons that
remain unclear \cite{bonn96b}) leads us to consider the opposite limit for
the strength of the scattering, the Born limit.

\begin{figure}
\begin{center}
\leavevmode
\epsfxsize=3.6in %% 
\epsfbox{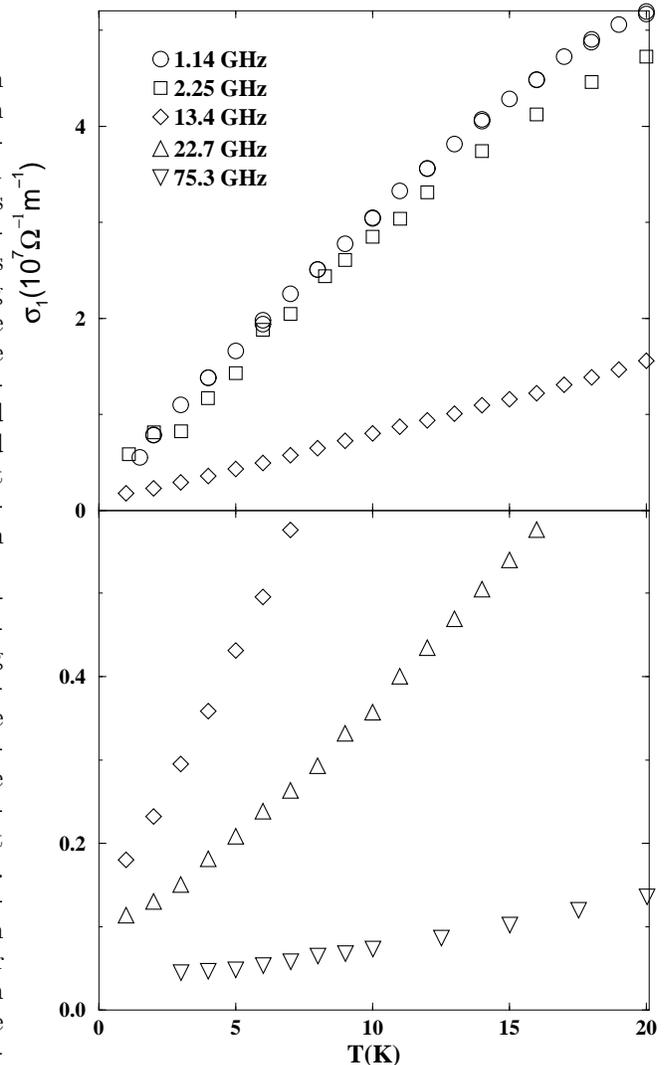}
\end{center}
\caption{This detailed view of the temperature dependence of the microwave
conductivity below 20 K shows a gradual evolution of the shape of
of $\sigma_1(T)$, from concave up at high frequencies to concave down at
the lowest frequencies. The quite linear temperature dependence seen here 
at 13 GHz seems to be an intermediate behaviour.}
\label{fig:sigmalowT}
\end{figure}

For impurity scattering 
in the Born limit, the crossover energy scale is exponentially small, so one
does not necessarily expect to see the universal conductivity limit
until measurements are performed well below 1 Kelvin. In fact, in the 
microwave measurements presented here we are not necassarily at low
enough temperature {\it or frequency} to observe any simple limiting behaviour.
So, in this situation we compare our results qualitatively to numerical
calculations performed in the Born limit by Hirschfeld et al.
\cite{hirschfeld94}. They found that at very low frequency $\sigma_1(T)$
rises rapidly from the universal zero temperature limit to a much larger
conductivity that depends upon the impurity scattering rate. It 
then remains
fairly temperature independent until inelastic scattering processes become
important. At higher frequencies $\sigma_1(T)$ becomes smaller and moves through
a whole range of behaviours, varying from mostly sub-linear in T at low
frequency, through a quasi-linear temperature dependence at intermediate
frequencies, to a faster than linear temperature dependence at high frequencies.
Figure \ref{fig:sigmalowT} shows behaviour in the measured
microwave conductivity
that is similar to this Born limit result in some of its qualitative
features. 
The overall conductivity has a magnitude that varies with purity, from
the quite high values seen here, through to very low, flat conductivities
observed in Ni-doped samples \cite{bonn94}.
More importantly, here we
do clearly see for the first time that the linear behaviour
of $\sigma_1(T)$ is an intermediate behaviour, albeit one that
survives over a substantial range of frequency and temperature.
The evolution in the shape of $\sigma_1(T)$ at low T clearly
falls outside of our phenomenological model, which would have predicted a
linear temperature dependence at all of the frequencies shown in Fig. 
\ref{fig:sigmalowT}. That is, if $\sigma_1(\omega)$ were really Drude-shaped
with a temperature independent width, then $\sigma_1(T)$ would exhibit the 
same temperature dependence at all frequencies; namely, the linear
temperature dependence of the normal fluid density. 

At our lowest frequencies the data
does move towards sub-linear as expected in the Born limit.
However, one perhaps important
difference from the theoretical calculations is that the trend does
not seem to continue
below 2 GHz. The data stops evolving towards the expected low frequency
behaviour, which is a rapid leap upwards to a
constant value. The reason for this is not yet clear, so we are left for the
moment with certain qualitative features that seem in accord with Born limit
scattering. Other features of the data seem to echo the expectations of a 
conductivity spectrum with a Drude-like lineshape that has a temperature
independent width; namely, the similarity in the 1 and 2 GHz
curves and the relatively large
frequency and temperature range over which $\sigma_1(T)$ seems to be
linear in T. It is the latter aspects of the data that lead to the 
phenomenological model being a fairly good description of the conductivity
spectra.

\begin{figure}
\begin{center}
\leavevmode
\epsfxsize=3in %% 
\epsfysize=2.98in %% 
\epsfbox{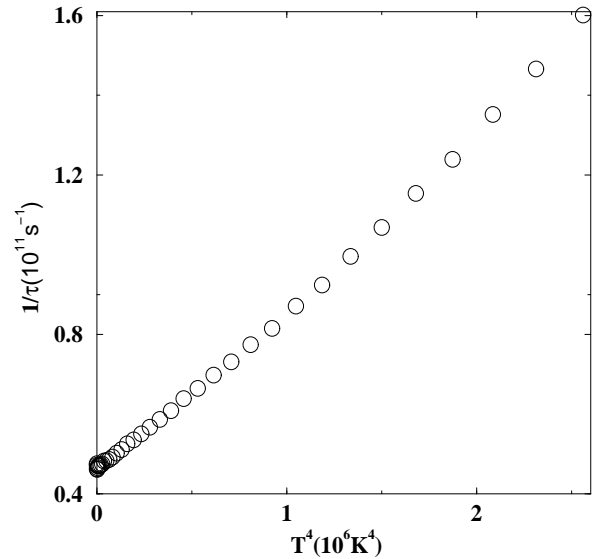}
\end{center}
\caption{When the temperature dependence of the scattering rate of thermally
excited quasiparticles is plotted
against $T^4$ as shown above it is clear that the onset of inelastic
scattering is very rapid, at least $T^4$ and even faster above 30 K.}
\label{fig:tau2}
\end{figure}

Although the foregoing discussion indicates that the $\sigma_1(\omega)$
spectra are not quite Drude shaped, there is still a characteristic width to the
peaks that is reasonably well parameterized by the Drude fits. Thus, the
plot of scattering rates shown in Fig. \ref{fig:tau1} still provides a
reasonable measure of the narrowness of the
peak at low temperature and its rapid
broadening above 20 K.
In Fig. \ref{fig:tau2} we plot this width, which we identify with the
quasiparticle
scattering rate, versus $T^4$. Just above 20 K, the
initial onset of this increase in scattering
appears to be at least as rapid as $T^4$ and rises even more quickly at 
higher temperatures. This type of rapid temperature dependence of the 
quasiparticle scattering time would be expected in any situation where
the inelatic scattering comes from interactions that become gapped below
$T_c$. A number of early calculations tackled the problem of the collapse
of the scattering rate below $T_c$ in this way. Early on, Nuss et al.
explained the peak in their THz conductivity measurements
in this manner \cite{nuss91}.
Littlewood and
Varma studied this type of effect in a marginal
Fermi liquid \cite{littlewood92}, the idea being that below $T_c$ a gap opens
up in the scattering spectrum. Statt and Griffin similarly studied the
effect of the opening of a gap in the spectrum of spin fluctuations 
\cite{statt92}. All of this work predated the solid identification of the 
$d_{x^2-y^2}$ pairing state, so isotropic $s$-wave gaps were assumed in 
the calculations. Quinlan et al. studied a model of quasiparticle 
scattering in which the lifetimes were associated with spin fluctuation
scattering. They studied the effects of both $s-$wave and $d-$wave
gaps opening up in the spin fluctuation spectrum \cite{quinlan94}. Since
we now know that the gap in this material has $d_{x^2-y^2}$ symmetry, this
latter calculation is most directly relevant to our measurements here. 
In particular they found
that at temperatures well below $T_c$, the quasiparticle
lifetime increases as $T^3$ and even faster than this as $T_c$ is approached.
In a comparable temperature range, the scattering rate that we extract from
the width of $\sigma_1(\omega)$ is closer to $T^4$. 
Thus, the temperature dependence of the inelastic scattering rate seems to be
about one power of T faster than the lifetime calculations based on a
gapping of the spin fluctuation spectrum. 

\section{Conclusions}
We have presented here the most complete microwave measurements yet obtained on
a crystal of a high temperature superconductor. These spectra
reveal a wealth of detail regarding the conductivity spectrum $\sigma_1(\omega)$
of the thermally excited quaiparticles in the superconducting state. We find
that $\sigma_1(\omega)$ has a Drude-like shape. Although in detail
there are 
deviations from this shape, there is nevertheless a well defined
characteristic width
which can be associated with the scattering time of the quasiparticles. Below 
$T_c$ the width collapses rapidly and it becomes easily discernible in the 
microwave spectral range at temperatures below 55 K. In the range between
55 K and 20 K we find that the collapse of the scattering rate varies at least
as fast as $T^4$. By 20 K the width becomes extremely narrow and nearly 
temperature independent. This narrow width of only 8 GHz corresponds to a 
mean free path as high as 4 $\mu m$ if we interpret the width as being 
a direct
measure of the elastic scattering rate due to impurities. We find that some
features of $\sigma_1(\omega,T)$ below 20 K are in accord with quasiparticle
scattering in the Born limit for a $d_{x^2-y^2}$ superconductor, in particular,
a gradual evolution of the shape of $\sigma_1(T)$ from sublinear T dependence
at low $\omega$, to quasilinear and then faster at higher $\omega$. However,
there remain
discrepancies that might best be settled by a detailed 
numerical calculation aimed at fitting the observations presented here.
Such fits must come to grips with the observation that the behaviour of
$\sigma_1(\omega,T)$ below 20 K is reasonably well described by a model
involving quasiparticle scattering with a temperature-independent lifetime.

\acknowledgments

We are greatly indebted to A.J. Berlinsky and C. Kallin for many helpful
discussions regarding transport properties in d-wave superconductors.
We wish also to acknowledge helpful conversations with
D.J. Scalapino, P.J. Hirschfeld, P. A. Lee
and G. Sawatsky, and are grateful for the opportunity to carry out some of this 
work at the Aspen Center for Physics.
We also
thank J. Trodahl for his contribution to the design of the 
75 GHz cavity setup.
This research was supported by the Natural Science and Engineering 
Research Council of Canada and the Canadian Institute for Advanced
Research. DAB acknowledges support from the Sloan
Foundation.

\end{document}